\documentclass[aps,prl,reprint,superscriptaddress]{revtex4-1}
\pdfoutput=1

\usepackage{graphicx,amssymb,amsmath,xcolor}
\usepackage[version=4]{mhchem}
\usepackage{layouts}
\usepackage{inputenc}
\begin{document}

\title{Exploring the Absorption Spectrum of Simulated Water from MHz to the Infrared}
\author{Shane Carlson}
\author{Florian N. Br{\"u}nig}
\author{Philip Loche}
\affiliation{Fachbereich Physik, Freie Universit{\"a}t Berlin, Arnimallee 14, 14195 Berlin, Germany}
\author{Douwe Jan Bonthuis}
\affiliation{Institute of Theoretical and Computational Physics, Graz University of Technology, 8010 Graz, Austria}
\author{Roland R. Netz}
\affiliation{Fachbereich Physik, Freie Universit{\"a}t Berlin, Arnimallee 14, 14195 Berlin, Germany}
\date{February 29, 2020}
\pacs{}

\begin{abstract}
Absorption spectra of liquid water at 300 K are calculated from both classical and density functional theory molecular dynamics simulation data, which together span from 1 MHz to hundreds of THz, agreeing well with experimental data qualitativley and quantitavely over the entire range, including the IR modes, the microwave peak, and the intermediate THz bands. The spectra are decomposed into single-molecular and collective components as well as into components due to molecular reorientations and changes in induced intramolecular dipole moments. These decompositions shed new light on the motions underlying the librational and translational (hydrogen-bond stretch) bands at 20 and 5 THz respectively:  interactions between donor protons and acceptor lone pair electrons are shown to be important for the line shape in both librational and translational regimes, and in- and out-of-phase librational dimer modes are observed and explored.
\end{abstract}
\maketitle

In spectroscopy of neat liquid water, there are several features appearing in simulated and experimental absorbtion spectra from the MHz regime up to hundreds of THz. Broadly speaking, five of the most conspicuous of these features are absorption bands at roughy 100, 50, 20, and 5 THz, and 20 GHz.
The bands at 100 and 50 THz are attributed to the OH-stretch and OH-bend modes of individual water molecules respectively \cite{1963_Buijs_JCP, 1964_Walrafen_JCP, 1990_Walfren_JPC, 2008_Auer_JCP}. 
The bands at 20 and 5 THz have long been attributed to molecular librations and translations respectively, i.e.\ hindered rotational and translational oscillations of an entire molecule within the HB (hydrogen-bond) network \cite{1964_Walrafen_JCP}. This view has since been confirmed by further experimental and MD (molecular dynamics) simulation studies \cite{2001_Keutsch_PNAS, 2010_Debnath_JCP, 2017_Chakraborty_ACR}.
Regarding the broad band near 20 GHz, which closely follows the characteristic form of a Debye function, though modified on the high-frequency side, there is less agreement or understanding. Classically, it was treated simply in the context of the Debye-Stokes-Einstein relation for free rotational diffusion of polar liquids \cite{book_1929_Debye_Polar, 1997_Ronne_JCP}. Later, simulations of water revealed molecular reorientations occuring abruptly upon the exchange of hydrogen-bond partners, contravening the free diffusion picture \cite{2006_Laage_sci}. These jump-like reorientations have also been contextualized as steps in the migration of Bjerrum-like defects through the HB network \cite{2016_Popov_PCCP, 2017_Elton_PCCP}. Markov-state modelling revealed HB-partner exchange processes to consist both of reorientations of lone molecules and of synchronous reorientations of multiple molecules, i.e.\ of an intricate combination of self and collective processes \cite{2018_Schulz_JCP}.

Calculating spectra from data of two complementary simulation methods--forcefield MD simulations of a classical rigid water model (SPC/E \cite{1987_Berendsen_JPC}) and DFT (density functional theory) MD simulations--we are able to produce a spectrum stretching from 1 MHz up to hundreds of THz that captures all five of these features, and agrees both qualitatively and quantitavely with experimental spectra without any rescaling of data. The two methods agree well with one another in the THz regime, and so might in principle be applied to a variety of systems to generate complementary spectra covering a very wide range of frequecies. Note that while we  discuss absorption spectra here, i.e.\ linear spectroscopy, many of the features we extract by decomposition can also be studied experimentally by 2D-IR spectroscopy \cite{2005_Eaves_PNAS, 2007_Auer_PNAS, 2008_Kraemer_PNAS, 2008_Yagasaki_JCP, 2016_Ito_JCP}.

\medskip

The band in water IR spectra peaking at $\sim$20 THz has long been attributed to librations, i.e.\ hindered partial rotations of water molecules in the HB network \cite{1964_Walrafen_JCP}. In agreement with Raman spectra of water, normal mode analysis of intermolecular motions in a $C_{2v}$-symmetric tetrahedral water cluster predicted two IR-active librational modes \cite{1964_Walrafen_JCP, inbook_1972_Walrafen_Physics, 1990_Walfren_JPC}. Two librational modes were also later inferred from experimental IR spectra, and a Carr-Parrinello simulation study found two collective librational modes, one negative and one positive \cite{1995_Zelsmann_JMS, 2008_Chen_PRB}. We find the collective librational modes  to arise from distinct in- and out-of-phase dimer modes, i.e.\ modes where two molecules that share a HB librate either synchronously or antisychronously. Further, we decompose the spectrum into components due to reorientations  of the mean molecular dipole (``orientational'') and induced intramolecular dipole moments (``induced''), and demonstrate the existence of a  dipole moment induced upon libration via the attraction of acceptor lone pair electrons to donor hydrogens. This dipole moment is induced counter to the libration of the molecule and attenuates the libration peak significantly at 20 THz.

\begin{figure*}[!ht]
	\centering
	\includegraphics{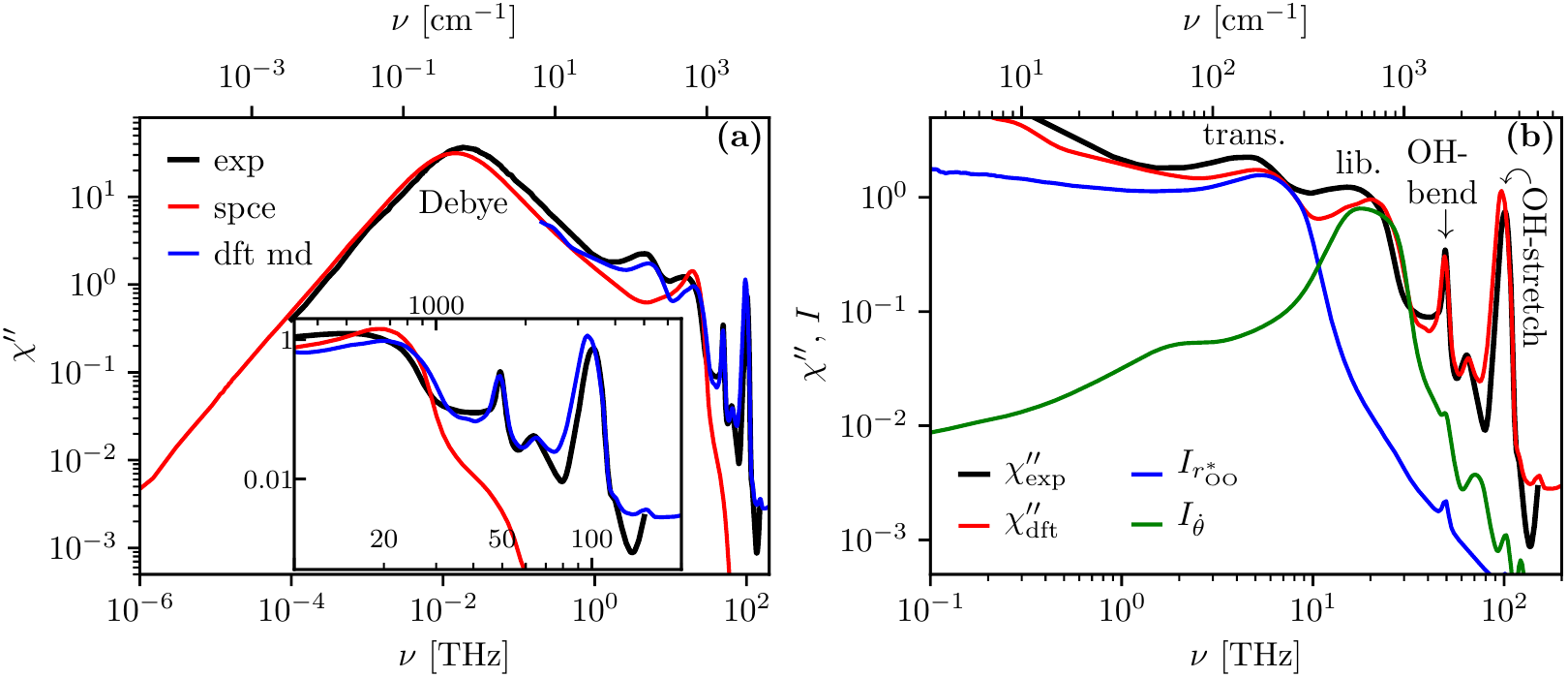}
	\caption{\textbf{(a)} A comparison of spectra from SPC/E and DFT MD simulations of bulk water at 300 K with experimental data. The same data over the highest frequencies is shown in the inset. \textbf{(b)} Power spectra for two physically relevant coordinates from the DFT MD simulation, namely $(i)$ the oxygen-oxygen distance for each pair of H-bonded molecules, $r_\mathrm{OO}^*$, and $(ii)$ the instantaneous angular velocity of the dipolar orientation of individual molecules, $\dot{\theta}$. Both power spectra are rescaled here for comparison.}
	\label{fig:fig1}
\end{figure*}

The band visible at $\sim$5 THz is attributed to hindered translations of a molecule in the HB network, i.e.\ HB stretching modes \cite{1964_Walrafen_JCP, 1990_Walfren_JPC, 1996_Walrafen_JPC, 1999_Ohmine_ACR, 1997_Silvestrelli_CPL, 2010_Heyden_PNAS}. It is conspicuous in both IR and Raman spectra of water. In multiple simulation studies, the IR activity of the translation band has been attributed to intermolecular charge transfer among H-bonded molecules \cite{2005_Sharma_PRL, 2011_Torii_JPCB, 2018_Sidler_JCP}. In spectra of classical forcefield simulations, this band was absent, including for rigid, flexible, and even polarizeable models \cite{2011_Torii_JPCB, 2012_Heyden_JPCL, 2015_Sega_JCPA}, until it was reproduced by including charge transfer among H-bonded molecules in post processing \cite{2018_Sidler_JCP}. Our aforementioned orientational/induced decomposition paints a picture where it is induced intramolecular dipole moments, specifically those due to interactions between lone pair electrons and donor hydrogens, that largely underlie the IR activity in the translation band -- a picture that is consistent with charge transfer.

\medskip

In summary, using a combination of forcefield and ab initio simulation methods, we generate liquid water spectra that cover over eight orders of magnitude, from 1 MHz to hundreds of THz, agreeing well with published experimental spectra qualitatively and quantitavely, including all five major features appearing in experimental dielectric, THz and IR spectra of water. The combination of self/collective and orientational/induced decompositions proves a simple means of probing spectral features in simulated spectra of liquid water; we use them to understand the motions underlying the translation and libration bands.

\section{Results and Discussion}

The frequency-depenedent complex electric susceptibility $\chi(\nu)$ is given by the fluctuation dissipation relation for the total system polarization $\boldsymbol{P}(t)$,
\begin{equation}
	\chi(\nu) = \frac{-1}{3 V \varepsilon_0 k_B T } \displaystyle\int_0^\infty \mathrm{d}t \;
	e^{-2\pi i \nu t} \frac{d}{dt}\langle \boldsymbol{P}(0) \cdot \boldsymbol{P}(t) \rangle \,,
	\label{eq:FDT_for_P}
\end{equation}
where $V$ is the system volume (see SI Section S1). Eq.\ \eqref{eq:FDT_for_P} corresponds to the classical limit; quantum correction schemes exist but are not employed here \cite{2004_Ramirez_JCP}. From Eq.\ \eqref{eq:FDT_for_P}, the Wiener-Khinchin theorem yields an expression for the dissipative imaginary part of $\chi(\nu)$,
\begin{equation}
\chi^{\prime\prime}(\nu) = \frac{\pi}{3 L_t V\varepsilon_0 k_B T } \;\nu \; \left\lvert \widetilde{\boldsymbol{P}}(\nu) \right\rvert^2 \,,
\label{eq:cpp_implement}
\end{equation}	
where $ L_t $ is the length in time of $\boldsymbol{P}(t)$, and $\widetilde{\boldsymbol{P}}(\nu)$ is its Fourier transform (see SI Section S2).
We implement Eq.\ \eqref{eq:cpp_implement} to extract spectra from $NVT$ MD simulations of liquid water at 300 K, of both classical MD (SPC/E, and in the SI Section S3, TIP4P/2005f \cite{2011_Gonzalez_JCP}) and DFT MD type. For the DFT MD trajectory, Wannier centers are calculated in order to obtain the polarization. More details are given in the Methods section.

\begin{figure*}
	\includegraphics{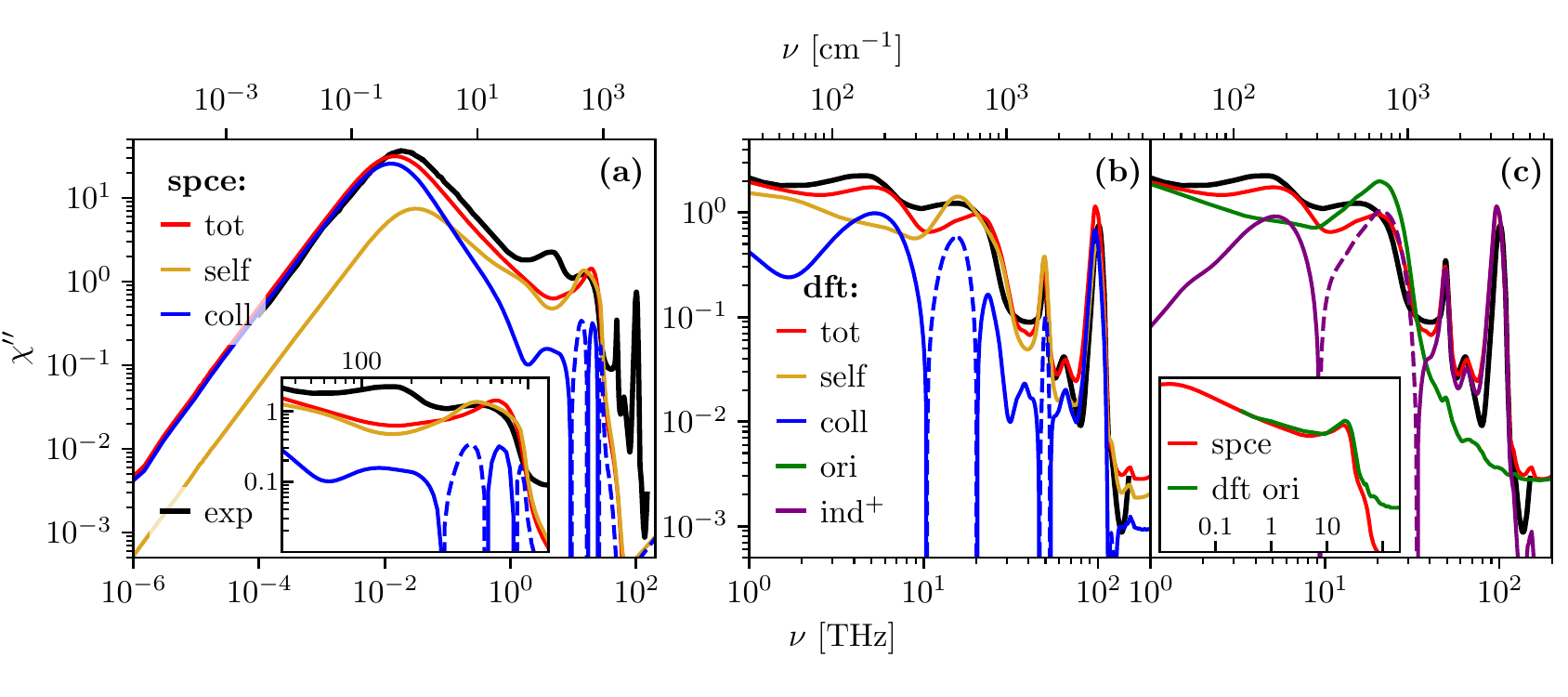}
\caption{Simulated spectra plotted with their respective decompositions. The heavy black curve represents experimental data. Dashed portions of spectral components represent negative data. \textbf{(a)} SPC/E and \textbf{(b)} DFT MD spectra decomposed into their self and collective components accordin to Eqs.\ \eqref{eq:P_SC_decomp} and \eqref{eq:chi_SC_decomp}. \textbf{(c)} DFT MD spectrum decomposed into orientational and induced components according to Eq.\ \eqref{eq:chi_oip_decomp}. \textbf{Inset:} The orientational component of the DFT MD spectrum compared with the full SPC/E spectrum in log-log.}
\label{fig:basic_decomp_triplot}
\end{figure*}

Figure \ref{fig:fig1}a shows a direct comparison of the  DFT MD and SPC/E spectra with a compiled experimental spectrum of water at $300 \pm 2$ K (see Methods section). Error estimates are calculated for all spectral data plotted, but are omitted from plots for clarity; an example plot with error estimates is shown in Section S4 of the SI. No simulated $\chi^{\prime\prime}$ data in this work are fit or rescaled in any way following the implementation of Eq.\ \eqref{eq:cpp_implement}, thus Figure \ref{fig:fig1}a represents a direct comparison of simulated and experimental spectra. The SPC/E spectrum extends down to 1 MHz, agreeing rather well with the experimental data from the lowest experimental data point at 100 MHz, up until $\sim$1 THz, reproducing the Debye peak magnitude and position, albeit with a small red-shift. The SPC/E spectrum does not capture the translation band at 5 THz, but does show a strong signature in the libration band at $\sim$20 THz which is expected for a model where polarization only changes due to molecular reorientations. The DFT MD simulation is computationally very costly, which limits the trajectory length (ours is 200 ps) and restricts the spectrum to higher frequencies. The DFT MD spectrum does however closely reproduce all gross features. The DFT MD and classical spectra agree well near 0.1 THz, a region where no modes involving induced intramolecular dipole moments are expected -- this lends credibility to the methods. Thus, we can produce spectra that compare directly to experiment, both qualitatively and quantitively, from 1 MHz well into the THz regime.

Figure \ref{fig:fig1}b begins an investigation into the translation and libration bands: it shows the experimental and total DFT MD spectra, along with power spectra, defined as $I_x(\nu) \sim | \tilde{x}(\nu) |^2$, of two physically relevant quantities calculated from the DFT MD trajectory, normalized for comparison to the susceptibility curves. The first is of $x=r_\mathrm{OO}^*$, the oxygen-oxygen distance between H-bonded molecules (HBs are defined according to the geometrical criterion defined in Ref.\ \cite{1996_Luzar_PRL}), whose power spectrum peaks very strongly at $\sim$5 THz, matching rather well with the position of the translation band. The second power spectrum is of $x=\dot{\theta}$, the angular velocity of the mean dipolar orientation of individual molecules, whose power spectrum peaks strongly at $\sim$20 THz, matching the position of the libration band. These power spectra confirm that \emph{motions} corresponding to HB stretching and molecular librations occur at 6 and 20 THz respectively; it remains to examine whether and how these motions result in absorption at the corresponding frequencies.

\textbf{Self/Collective Decomposition:} The susceptibility is decomposed into two components: that due to autocorrelations in the dipole moments of single molecules, and that due to cross correlations among the dipole moments of different molecules \cite{2008_Chen_PRB, 2010_Heyden_PNAS}. To this end, the system polarization is written as the sum over the $N$ molecular dipole moments,
$\boldsymbol{P}(t) =  \sum_{i=1}^N \boldsymbol{p}^{i}(t)$,
where $\boldsymbol{p}^{i}(t)$ is the dipole moment for the $i^\mathrm{th}$ molecule.
Then the correlation function appearing in Eq.\ \eqref{eq:FDT_for_P} can be decomposed as
\begin{equation}
	 \langle \boldsymbol{P}(0) \cdot \boldsymbol{P}(t) \rangle
= \sum_{i=1}^N \left\langle \boldsymbol{p}^{i}(0)\cdot\boldsymbol{p}^{i}(t)\right\rangle + \sum_{i=1}^N  \left\langle \boldsymbol{p}^{i}(0) \cdot  \displaystyle \sum_{ j\neq i} \boldsymbol{p}^{j}(t)  \right\rangle\,,
	 \label{eq:P_SC_decomp}
\end{equation}
giving
\begin{equation}
\chi(\nu) = \chi_\mathrm{self}(\nu) + \chi_\mathrm{coll}(\nu)\,.
	 \label{eq:chi_SC_decomp}
\end{equation}

Figures \ref{fig:basic_decomp_triplot}a and b show decompositions of simulated spectra alongside experimental data. The dashed portions of spectral component curves represent negative contributions arising from anticorrelations. The SPC/E spectrum shown in Figure \ref{fig:basic_decomp_triplot}a becomes very collective in the low-frequency regime. The collective part contains a clear signature of the translation band near 5 THz, and negative/positive splitting in the libration band, the signature of out-of-phase and in-phase librations of H-bonded molecules, as further discussed below. 

The self/collective decomposition of the DFT MD spectrum shown in Figure \ref{fig:basic_decomp_triplot}b is in broad agreement with that shown in Ref.\ \cite{2008_Chen_PRB}. It reveals the highly collective nature of the translation band and positive and negative modes in the collective part of the libration peak. The libration peak as a whole is revealed to be predominantly single-molecular. The signature of the OH-bend at $\sim$50 THz is strongly single-molecular with a small negative collective contribution, while the OH-stretch is equal parts collective and single-molecular.

\begin{figure}
\includegraphics{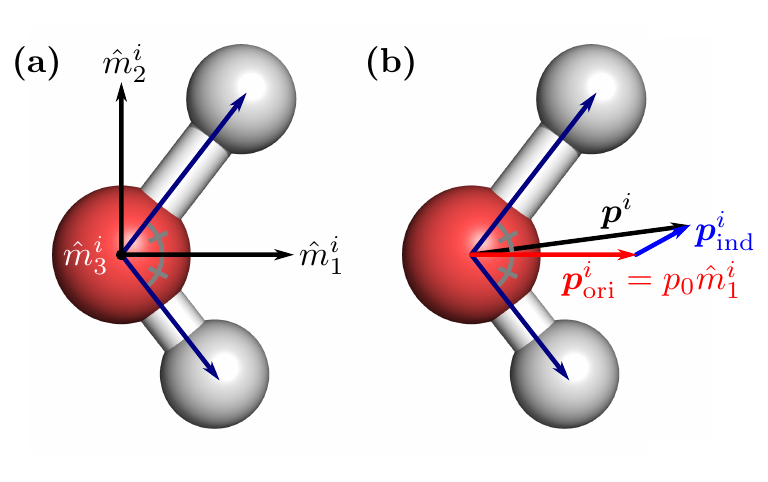}
\caption{\textbf{(a)} Orientational molecular coordinates. OH bonds are shown with different lengths to emphasize that $\hat m_1^i$ depends only on the respective directions of $\boldsymbol{r}_{\mathrm{OH}^1}^i$ and $\boldsymbol{r}_{\mathrm{OH}^2}^i$, shown as dark blue vectors. \textbf{(b)} Orientational and induced components of the molecular dipole. }
\label{fig:2mol_m}
\end{figure}

\textbf{Orientational/Induced Decomposition:} We devise a further decomposition scheme where the dipole moment of each molecule is split into two parts, That due to the mean dipole, which changes only under reorientations of the molecule, and that due to induced dipole moments. An orientational axis is defined for each molecule at each timestep as the bisector of $\angle$HOH, parametrized for the $i^\mathrm{th}$ molecule via the unit vector $\hat m_1^i(t)$.
This choice makes the orientation independent of OH-distance and $\angle$HOH. Additionally, we define two more unit vectors: $\hat m_2^i(t)$ lies in the water plane and orthogonal to $\hat m_1^i(t)$, and $\hat m_3^i(t)$ is orthogonal to the water plane. Thus, $(\hat m_1^i(t),\, \hat m_2^i(t),\, \hat m_3^i(t))$ form an orthonormal basis determined by the molecular orientation, as illustrated in Figure \ref{fig:2mol_m}a.

The molecular dipole $\boldsymbol{p}^{i}$ is decomposed into two components which we term ``orientational'' and ``induced'', 
as  illustrated in  Figure \ref{fig:2mol_m}b. The orientational component of the $i^\mathrm{th}$ molecule is defined as a dipole moment of fixed magnitude parallel to $\hat m_1^i$,
\begin{equation}
\boldsymbol{p}_{\mathrm{ori}}^i(t) \equiv p_0 \, \hat m_1^i(t) \,,
\label{eq:p_ori}
\end{equation}
where the magnitude $p_0$ is the average of the dipole moment $\boldsymbol{p}^i(t)$ projected onto $\hat m_1^i(t)$, found to be $p_0 = 0.615$ e{\AA} in the DFT MD trajectory (cf.\ $p_0=0.489$ e{\AA} for SPC/E).

The second component is the remaining portion of the molecular dipole moment, which we term the \emph{induced} component;
\begin{equation}
\boldsymbol{p}_\mathrm{ind}^i(t) \equiv \boldsymbol{p}^{i}(t) - \boldsymbol{p}_\mathrm{ori}^i(t) \,.
\label{eq:p_ind}
\end{equation}
The definitions \eqref{eq:p_ori} and \eqref{eq:p_ind} ensure that the induced dipole moment has zero mean. These molecular components can be summed to obtain the system polarization components,
\begin{equation}
	\boldsymbol{P}_\mathrm{ori}(t) = \displaystyle \sum_{i=1}^N \boldsymbol{p}_\mathrm{ori}^i(t)\,,
	\hspace{5mm}
	\boldsymbol{P}_\mathrm{ind}(t) = \displaystyle \sum_{i=1}^N \boldsymbol{p}_\mathrm{ind}^i(t)\,,
\label{eq:ps_resummed}
\end{equation}
which  yields three susceptibilities, namely those  due to $\left\langle \boldsymbol{P}_\mathrm{ori}(0) \cdot \boldsymbol{P}_\mathrm{ori}(t)\right\rangle$, $\left\langle \boldsymbol{P}_\mathrm{ind}(0)\cdot \boldsymbol{P}_\mathrm{ind}(t) \right\rangle$ and $\left\langle \boldsymbol{P}_\mathrm{ori}(0)\cdot \boldsymbol{P}_\mathrm{ind}(t) \right\rangle$, which gives the full decomposition
$\chi(\nu) = \chi_\mathrm{ori}(\nu) + \chi_\mathrm{ind}(\nu) + \chi_\mathrm{ori \times ind}(\nu)$.
In the interest of separating components into those contingent on induced dipole moments and those not, we define further
$\chi_{\mathrm{ind}^+}(\nu)  \equiv \chi_\mathrm{ind}(\nu) + \chi_\mathrm{ori \times ind}(\nu)$,
giving the bipartite decomposition
\begin{equation}
	\chi(\nu) =   \chi_\mathrm{ori}(\nu) + \chi_{\mathrm{ind}^+}(\nu)\,.
\label{eq:chi_oip_decomp}
\end{equation}

The DFT MD spectrum decomposed according to Eq.\ \eqref{eq:chi_oip_decomp} is shown in Figure \ref{fig:basic_decomp_triplot}c. The translation band and IR peaks are apparently contingent on induced dipole moments, while the libration band is massively orientational. The inset compares the DFT MD orientational component to the full SPC/E spectrum, revealing close agreement in shape and magnitude below about 50 THz, indicating a close similarity in the gross dynamics of whole molecules between rigid classical and AIMD models of water. This implies that what sets AIMD spectra apart from those generated from classical models is, in the main, changes in induced dipole moments, which incidentally may be added in post processing to obtain spectral features, as has been done e.g.\ in the form of charge transfer \cite{2018_Sidler_JCP}.

\textbf{A Closer Look at Translations and Librations:} Figure \ref{fig:basic_decomp_triplot}b shows the translation band near 5 THz to be largely collective (cf.\ \cite{2008_Chen_PRB, 2010_Heyden_PNAS}), which implies that the peak arises from correlated changes in dipole moments among different molecules. Figure \ref{fig:basic_decomp_triplot}c shows it to be contingent on induced dipole moments, i.e.\ involving induced$\times$induced and induced$\times$orientational correlations, which is not in general inconsistent with the partial charge transfer picture of Refs.\ \cite{2005_Sharma_PRL, 2011_Torii_JPCB, 2018_Sidler_JCP}. In the calculation of Wannier centers, a small partial charge transferred across a HB  from acceptor to donor is relocalized at a lone-pair Wannier center near the acceptor, manifesting in a spatial shift of the Wannier center and inducing a dipole moment in the acceptor molecule (note that the simple assignment of Wannier centers to the nearest oxygen always resulted in four Wannier centers per molecule, i.e.\ neutral molecules). Indeed, it is shown in Section S5 of the SI that lone-pair Wannier center positions are highly dependent on nearby attractive donor hydrogens. In our picture then, the activity at $\sim$5 THz is due to the attractive Coulomb force of donor hydrogens moving lone-pair electrons about with respect to their parent molecules.

\begin{figure*}
\centering
\includegraphics{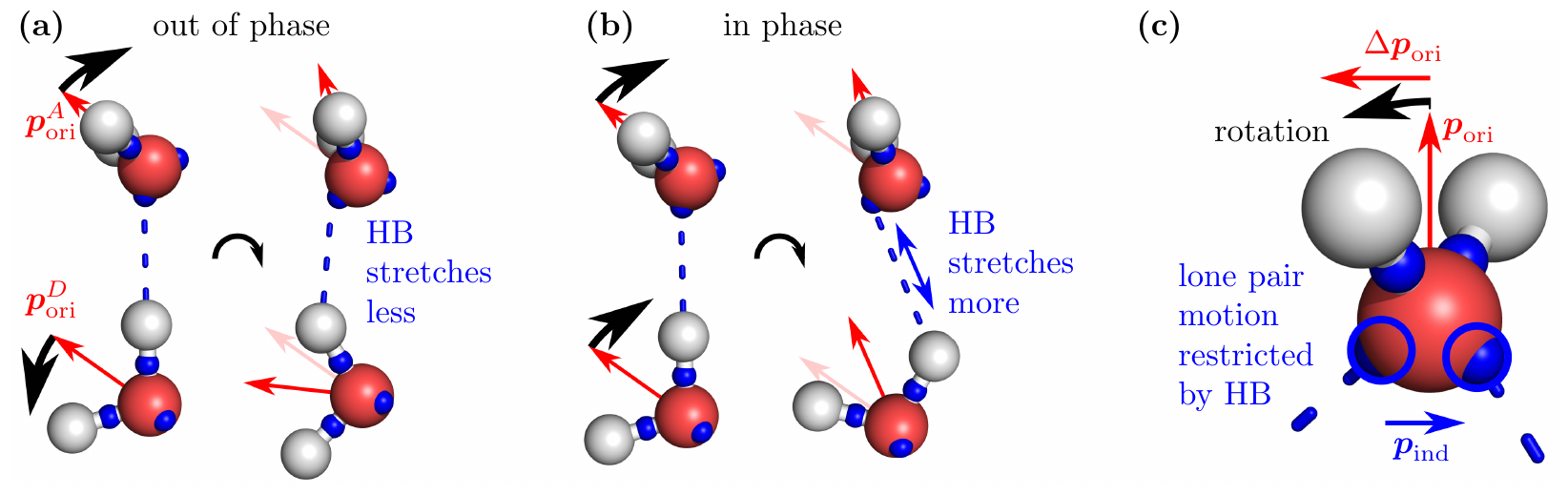}
\caption{Molecular schematics: small blue spheres indicate Wannier center positions and dashed blue lines indicate HBs. \textbf{(a)} Out-of-phase dimer libration mode. The shared HB is stretched less than for the in-phase mode.  \textbf{(b)} In-phase dimer libration mode. The shared HB is stretched further, resulting in a higher frequency. \textbf{(c)} A schematic illustrating how attractive donor hydrogens prevent lone pair electrons from freely rotating with their librating parent molecule.}
\label{fig:schematics_librations}
\end{figure*}

Figures \ref{fig:basic_decomp_triplot}a and b, representing the SPC/E and DFT MD methods, both show the libration band near 20 THz to be largely due to the self component and the collective spectrum to consist of  negative  and positive modes, which were observed previously \cite{2008_Chen_PRB}. These are due to  correlations within H-bonded dimers consisting of out-of-phase and in-phase dimer libration modes at $\sim$25 and $\sim$15 THz respectively, as illustrated in Figures \ref{fig:schematics_librations}a and b. The positive and negative modes appear for any combination of donor and acceptor librations about $\hat{m}_2^i$ and $\hat{m}_3^i$, and the configurations shown in Figures \ref{fig:schematics_librations}a and b only represent one choice of relative molecular orientations and librational axes. Thus the differences in frequency arise due to the shared HB in the dimer stretching more for in-phase motion, resulting in a greater restoring force and a higher frequency (see Section S6 of the SI for details).

Figure \ref{fig:basic_decomp_triplot}c shows the librational component to be, unsurprisingly, highly orientational. The induced$^+$ component is highly negative there, which is due to the orientational-induced cross component (shown in the SI Section S6.1). This negative correlation implies that a dipole moment  $ \boldsymbol{p}_\mathrm{ind}$ is induced opposite to the change in the mean dipole $\Delta \boldsymbol{p}_\mathrm{ori}$. This is due to the attraction of lone pair electrons by donor hydrogens; seen in the molecular frame of reference, these electrons rotate counter to the libration, inducing an opposite dipole moment, as illustrated in Figure \ref{fig:schematics_librations}c. This compensatory behavior explains why the libration peak for rigid classical forcefield water models, which lack electronic degrees of freedom, is more sharply peaked than experimental spectra (see Figure \ref{fig:basic_decomp_triplot}a). Indeed, the inset of Figure \ref{fig:basic_decomp_triplot}c shows the SPC/E spectrum and the orientational component of the DFT MD both taking on this characteristic shape at $\sim$20 THz.

\section{Conclusion and Outlook}

Using a combination of classical forcefield and ab initio MD simulations, we generate spectra for liquid water that agree well with experiment in frequency and magnitude from 1 MHz up to hundreds of THz. We interpret the spectra via a combination of decompositions, employing self/collective decompositions and a decomposition of the molecular dipole moment into orientational and induced components to unravel the molecular mechanisms underlying all bands. The translation band is shown to be contingent on induced molecular dipoles, the libration band, on molecular reorientations; the existence of librational dimer modes is inferred, and the role of intermolecular lone-pair/hydrogen interactions is elucidated. These spectral extraction and decomposition methods might be extended to a variety of liquids, especially polar liquids, but are immediately applicable to the analysis of liquid water in a wide variety of systems. For example, for hydration water in aqueous solutions and proteins, spectra can be decomposed according to hydration shell and correlated with microscopic motion.

\section{Methods}

\textbf{Simulation Methods:} One DFT MD and four classical forcefield MD simulations are carried out in the $NVT$ ensemble of 256 water molecules in bulk in a cubic box at 300 K. The box side length $L \simeq 2$ nm for 256 water molecules is large enough to prevent interactions with multiple periodic images in the forcefield simulations, where the Lennard-Jones interaction cutoff length is defined to be 0.9 nm. Temperature coupling for both DFT and forcefield simulations is carried out using the velocity-rescaling (CSVR) thermostat \cite{2007_Bussi_JCP}.

For the classical simulations, the rigid, 3-point water model SPC/E is simulated in GROMACS $2018.1$ using the leap-frog integrator \cite{1987_Berendsen_JPC, 1974_Hockney_JCP}. Starting configurations are drawn from an equilibrated $NpT$ simulation, for which the water density varies around $0.987 \pm 0.010$ g/cm$^3$.  Four simulations are carried out of varying length and write out frequency, 10 $\mu$s $@$ 1 ps,  1 $\mu$s $@$ 100 fs, 100 ns $@$ 10 fs, and 10 ns $@$ 0.5 fs, in order to cover a broad spectral range. 

The DFT MD simulation is carried out in CP2K 4.1 using a polarizable double-$\zeta$ basis set, optimized for small molecules and short ranges, for the valence electrons (DZVP-MOLOPT-SR-GTH), dual-space pseudopotentials, the BLYP exchange-correlation functional and D3 dispersion correction \cite{1992_Kendall_JCP, 2010_Grimm_JCP, 2014_Hutter_WIRCMS}. For a critical comparison of different DFT methods for water see Ref.\ \cite{2016_Gillan_JCP}. The cutoff for the plane-wave representation is optimized to 400 Ry. The time constant of the thermostat is set to 100 fs, which has been shown to be exceptionally good for preserving vibrational dynamics. The water density is 0.998 g/cm$^3$. The simulation trajectory is 200 ps in length with a time step of 0.5 femtoseconds. In post-processing, Wannier centers are calculated every 2 fs, and assigned to the molecule of the nearest oxygen, which always results in exactly four Wannier centers per water molecule. A charge of $-2e$ is assigned to each Wannier center, allowing for the calculation of the dipole moment.

\textbf{Calculation of Spectra:} Spectra are generated by dividing each simulation trajectory into ten equal segments, calculating $\chi^{\prime\prime}(\nu)$ for each segment, smoothing, and taking the mean, which allows for an estimation of the standard deviation. An implemention of Eq.\ \eqref{eq:cpp_implement} yields $\chi^{\prime\prime}(\nu)$ for a single segment, where $ \widetilde{\boldsymbol{P}}(\nu)$ is found by zero padding $\boldsymbol{P}(t)$ and applying a fast Fourier transform. Smoothing is carried out by convolution with a Gaussian function; the standard deviation $\sigma$ of the convoluting Gaussian are for the SPC/E spectra $10^{-4}$, $10^{-3}$, $10^{-2}$, $10^{-1}$ THz for the 10 $\mu$s, 1 $\mu$s, 100 ns, and 10 ns simulations respectively, and $1$ THz for the DFT MD spectra. Data are then thinned by removing all but a set of roughly log-spaced datapoints. The SPC/E spectra are convoluted with a Gaussian of $\sigma = 0.5$ THz a second time to lessen remaining noise.

\textbf{Experimental Data:} Experimental susceptibility datasets for liquid water near 300 K are compiled from several different sources, covering different frequency ranges to overlap. At a reported temperature of 25$^\circ$C are
Ref.\ \cite{1976_Schwan_JCP} 0.1 - 3 GHz,
Ref.\ \cite{1991_Barthel_BBPC} 1.7 - 89 GHz,
Ref.\ \cite{1989_Kaatze_JCED} 1.8 - 58 GHz, and
Ref.\ \cite{1990_Czumaj_MP} 57 - 315 GHz.
At a reported temperature of 27$^\circ$C is
Ref.\ \cite{1975_Downing_JGR} 0.3 - 150 THz.
Together, they cover the frequency range of 100 MHz to 150 THz. Refs.\ \cite{1976_Schwan_JCP, 1991_Barthel_BBPC, 1990_Czumaj_MP} are found via a compilation in Ref.\ \cite{2007_Ellison_JPCRD}. Section S7 of the SI includes more information, including a scatter plot of each dataset with error estimates shown.

\textbf{Acknowledgements:} We are grateful for the financial support of the Deutsche Forschungsgemeinschaft (DFG) via grants under SFB 1078 Project C1 and SFB 1114 Project C2.

\bibliography{spectroscopy}

\end{document}


\renewcommand{\Re}{\,\operatorname{Re}\:}
\renewcommand{\Im}{\,\operatorname{Im}\:}

\renewcommand{\thepage}{S\arabic{page}}
\renewcommand{\thesection}{S\arabic{section}}
\renewcommand{\thetable}{S\arabic{table}}
\renewcommand{\thefigure}{S\arabic{figure}}
\renewcommand{\theequation}{S\arabic{equation}}

\title{Exploring the Absorption Spectrum of Simulated Water from MHz to the Infrared
-
Supplementary Information}
\author[1]{Shane Carlson}
\author[1]{Florian N. Br{\"u}nig}
\author[1]{Philip Loche}
\author[2]{Douwe Jan Bonthuis}
\author[1]{Roland R. Netz}
\affil[1]{Fachbereich Physik, Freie Universit{\"a}t Berlin, Arnimallee 14, 14195 Berlin, Germany}
\affil[2]{Institute of Theoretical and Computational Physics, Graz University of Technology, 8010 Graz, Austria}
\date{February 29, 2020}

\maketitle

\section{Complex Electric Susceptibility from Fluctuation Dissipation Theorem}
\label{sec:SI_susc_FDT}

Assuming an isotropic medium perturbed by an applied electric field $\boldsymbol{E}$, whose Hamiltonian takes the form $H=H_0-\boldsymbol{P}\cdot\boldsymbol{E}$, the linear response of the total system dipole moment $\boldsymbol{P}$ (not to be confused with the polarization density) is described by the time-dependent electric susceptibility $\chi(t)$ via
\begin{equation}
	\boldsymbol{P}(t)  =   \displaystyle\int_{-\infty}^t \mathrm{d}t^\prime \;
	V \varepsilon_0 \chi(t-t^\prime)  \boldsymbol{E}(t^\prime)\,.
	\label{eq:SI_lin_resp_PE}
\end{equation}
where $V$ is the system volume. 
Fourier transforming Eq.\ \eqref{eq:SI_lin_resp_PE} gives
\begin{equation}
	\widetilde{\boldsymbol{P}}(\nu)  =  V \varepsilon_0  \chi(\nu) \widetilde{\boldsymbol{E}}(\nu) \,,
	\label{eq:SI_lin_resp_Fourier_PE}
\end{equation}
where $\chi(\nu)$ is the positive-domain Fourier transform of $\chi(t)$, and is known as the complex electric susceptibility, generalized electric susceptibility, or frequency-dependent electric susceptibility. $\chi(\nu)$ is a dimensionless, complex quanitity, denoted here as $\chi(\nu) = \chi^\prime(\nu) - i \chi^{\prime \prime} (\nu)$, in order that $\chi^{\prime \prime} (\nu)$ be positive for positive $\nu$.
The fluctuation dissipation relation expresses the linear response function $V \varepsilon_0  \chi(\nu)$ in terms of an equilibrium ensemble average 
\begin{equation}
	V \varepsilon_0 \chi(\nu) = \frac{-1}{3  k_B T} \displaystyle\int_0^\infty \mathrm{d}t \;
	e^{-2\pi i \nu t} \frac{\mathrm{d}}{\mathrm{d}t}\langle \boldsymbol{P}(0) \cdot \boldsymbol{P}(t) \rangle \,,
	\label{eq:SI_FDT_for_P}
\end{equation}
where the arithmetic mean has been taken over the three spatial dimensions.
Here we have taken the Fourier transform of the function $f(t)$ to be defined as
\begin{equation}
	\tilde{f}(\nu) = \int_{-\infty}^\infty \mathrm{d}t \; e^{-2\pi i \nu t} f(t)\,, 
	\label{eq:SI_ft}
\end{equation}
whose inverse Fourier-transform is
\begin{equation}
	f(t)  = \int_{-\infty}^\infty \mathrm{d}\nu \; e^{2\pi i \nu t} \tilde{f}(\nu)\,.
	\label{eq:SI_ift}
\end{equation}

\section{The Dissipative Part of the Susceptibility}
\label{sec:SI_dissipative_part}

Starting from Eq.\ \eqref{eq:SI_FDT_for_P} and using that $\langle \boldsymbol{P}(0) \cdot \boldsymbol{P}(t) \rangle$ is real, the imaginary (dissipative) part of $\chi(\nu)$ is found to be 
\begin{equation}
	\chi^{\prime\prime}(\nu) = \frac{-1}{3 V k_B T \varepsilon_0}  \displaystyle\int_0^\infty \mathrm{d}t \;
	\sin(2\pi \nu t) \;\frac{d}{dt}\,\langle \boldsymbol{P}(0) \cdot \boldsymbol{P}(t) \rangle \,.
	\label{eq:SI_cpp1}
\end{equation}
Because the autocorrelaton function is symmetric, the integrand is symmetric, thus
\begin{align}
	\chi^{\prime\prime}(\nu) &= \frac{-1}{3 V k_B T \varepsilon_0} \frac{1}{2} \displaystyle\int_{-\infty}^\infty \mathrm{d}t \;
	\sin (2\pi \nu t) \;\frac{d}{dt}\, \langle \boldsymbol{P}(0) \cdot \boldsymbol{P}(t) \rangle  \notag\\
	&= \frac{1}{3 V k_B T \varepsilon_0} \frac{1}{2}  \Im \displaystyle\int_{-\infty}^\infty \mathrm{d}t \;
	e^{-2\pi i \nu t}  \frac{d}{dt} \langle \boldsymbol{P}(0) \cdot \boldsymbol{P}(t) \rangle \notag\\
	&= \frac{\pi}{3 V k_B T \varepsilon_0}  \nu \Re \displaystyle\int_{-\infty}^\infty \mathrm{d}t \;
	e^{-2\pi i \nu t} \; \langle \boldsymbol{P}(0) \cdot \boldsymbol{P}(t) \rangle \,. \label{eq:cpp2}
\end{align}
Using that the Fourier transform of a symmetric function is always real-valued gives for the dissipative part,
\begin{equation}
	\chi^{\prime\prime}(\nu) = \frac{\pi}{3 V \varepsilon_0 k_B T} \,\nu\,\displaystyle\int_{-\infty}^\infty \mathrm{d}t \;
	e^{-2\pi i \nu t} \; \langle \boldsymbol{P}(0) \cdot \boldsymbol{P}(t) \rangle \,.\label{eq:cpp4}
\end{equation}
Application of the Wiener-Khinchin theorem (Eq.\ \eqref{eq:SI_WK} below) gives
\begin{equation}
\chi^{\prime\prime}(\nu) = \frac{\pi}{3 L_t V\varepsilon_0 k_B T } \;\nu \; \left\lvert \widetilde{\boldsymbol{P}}(\nu) \right\rvert^2 \,,	
\label{eq:SI_cpp_implement}
\end{equation}	
where $L_t$ is the length in time of $\boldsymbol{P}(t)$. Eq.\ \eqref{eq:SI_cpp_implement} is the Equation implemented in this work as it involves only a single Fourier transform of each Cartesian component of $\boldsymbol{P}(t)$ and otherwise simple array operations.

\subsection*{Wiener-Khinchin Theorem}

Assuming the available data of a signal $f(t)$ is limited to a finite time interval $\left[ 0,\,L_t \right]$, we formally define $f(t)$ as being zero outside this interval. We define the autocorrelation function of $f(t)$ as the mean over the interval $I$ (of length $L_t-|t|$) where there is available data
\begin{equation}
	C(t) = \frac{1}{L_t-|t|} \int_I \mathrm{d}t^\prime \; f^*(t^\prime)\, f(t^\prime+t)\,,
	\label{eq:SI_corr}
\end{equation}
where  $f^*(t)$ is the complex conjugate of $f(t)$. If $f(t)$ is an observable in an equilibrium system, then by the ergodic theorem, $C(t)$ is the best estimate of the equilibrium ensemble average $\left\langle f^*(0) f(t) \right\rangle$. For $t\ge 0$, $I = \left[ 0,\,L_t-t \right]$ , and for $t\le 0$,  $I = \left[ |t|,\,L_t\right]$. The integrand $f^*(t^\prime) \, f(t^\prime+t)$ is always zero for $t^\prime$ outside of $I$, so in both cases, the integration bounds can be extended arbitrarily to give a generally applicable expression,
\begin{equation}
	C(t) = \frac{1}{L_t-|t|} \int_{-\infty}^{\infty} \mathrm{d}t^\prime \; f^*(t^\prime) \, f(t^\prime+t)\,.
	\label{eq:SI_corr2}
\end{equation}
This step in the proof demands that in practice, the signal $f(t)$ be zero padded: that is, zeros of length $L_t$ should be appended to the end of $f(t)$ before Fourier transforming. Substituting Eq.\ \eqref{eq:SI_ift} for $f^*(t^\prime)$ and $f(t^\prime+t)$ gives 
\begin{align}
	C(t) &= \frac{1}{L_t-|t|} \int_{-\infty}^{\infty} \mathrm{d}t^\prime \int_{-\infty}^\infty \mathrm{d}\nu \; e^{-2\pi i \nu t^\prime} \tilde{f}^*(\nu)\, \int_{-\infty}^\infty \mathrm{d}\mu \; e^{2\pi i \mu (t^\prime+t)} \tilde{f}(\mu)\,, \notag\\
	&= \frac{1}{L_t-|t|} \int_{-\infty}^\infty \mathrm{d}\nu \;  \tilde{f}^*(\nu)\, \int_{-\infty}^\infty \mathrm{d}\mu \; e^{2\pi i \mu t} \tilde{f}(\mu) 
	\int_{-\infty}^{\infty}\mathrm{d}t^\prime \; e^{2\pi i t^\prime(\mu - \nu)} \,,\notag\\
	&= \frac{1}{L_t-|t|} \int_{-\infty}^\infty \mathrm{d}\nu \;  \tilde{f}^*(\nu)\, \int_{-\infty}^\infty \mathrm{d}\mu \; e^{2\pi i \mu t} \tilde{f}(\mu)\,
	\delta(\mu - \nu)\,,\notag\\
	&= \frac{1}{L_t-|t|} \int_{-\infty}^\infty \mathrm{d}\nu \; e^{2\pi i \nu t}\,  \tilde{f}^*(\nu)\, \tilde{f}(\nu)\,. \label{eq:SI_WK_align}
\end{align}
Rearranging and Fourier transforming both sides gives
\begin{equation}
	\tilde{f}^*(\nu)\, \tilde{f}(\nu) = \int_{-\infty}^\infty \mathrm{d}t \; e^{-2\pi i \nu t} \left( L_t-|t| \right) C(t) \,.
	\label{eq:SI_WK_intermediate}
\end{equation}
We assume correlations are large for small $t$ and decay over time, so both terms of Eq.\ \eqref{eq:SI_WK_intermediate} should be dominated by the small-$|t|$ regime. In this regime, in the limit of large $L_t$, $|t|/ L_t \rightarrow 0$, so we may neglect the $|t|$ term. Additionally, for large $L_t$, $C(t) \rightarrow \left\langle f^*(0) f(t) \right\rangle$. Thus Eq.\ \eqref{eq:SI_WK_intermediate} may be rewritten as the Wiener-Khinchin theorem
\begin{equation}
	\int_{-\infty}^\infty \mathrm{d}t \; e^{-2\pi i \nu t} \left\langle f^*(0) f(t) \right\rangle = \frac{1}{L_t}
	\left\lvert \tilde{f}(\nu)  \right\rvert^2 \,.
	\label{eq:SI_WK}
\end{equation}

\section{Spectrum of Flexible Classical Water Model TIP4P/2005f}
\label{sec:SI_TIP4P2005f}

\begin{figure}[!ht]
	\centering
	\includegraphics{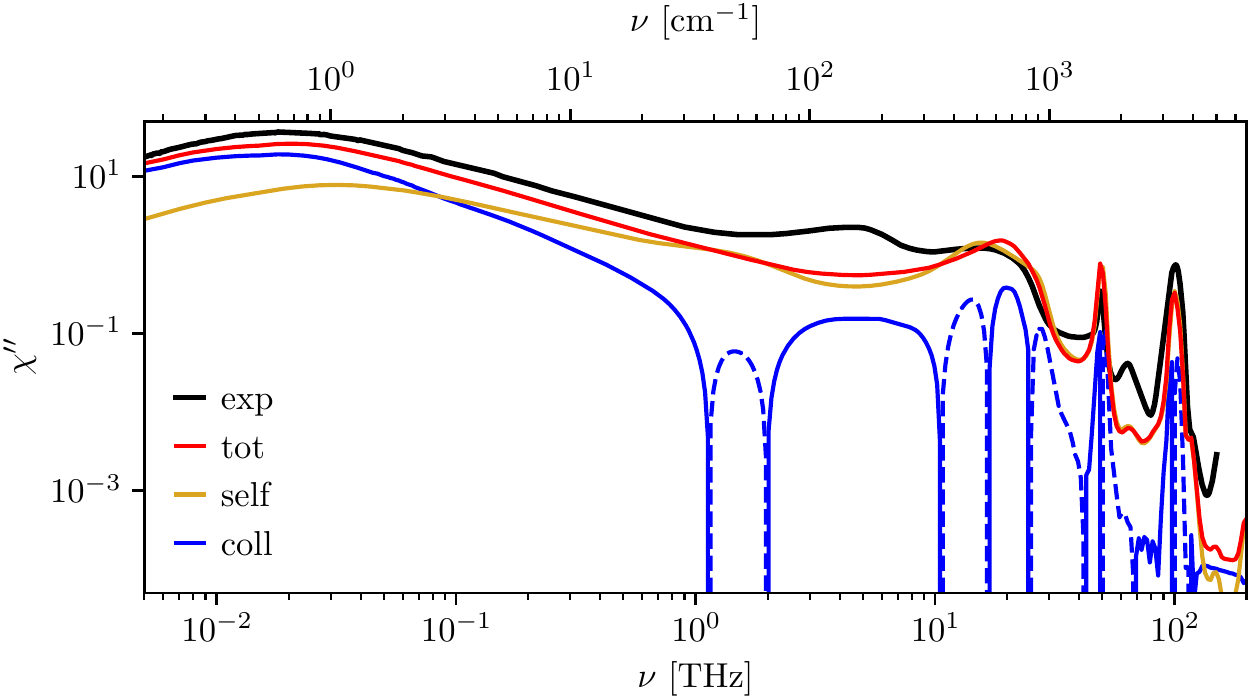}
	\caption{The TIP4P/2005f spectrum along with its self and collective components.}
	\label{fig:SI_tip4p2005f}
\end{figure}

We also calculated spectra using other classical water models, including of a 5 ns trajectory with a 0.5 fs writeout frequency of TIP4P/2005f, a flexible 4-point water model \cite{2011_Gonzalez_JCP}. The resulting spectrum, along with self and collective components, is shown in Figure \ref{fig:SI_tip4p2005f}. Here, smoothing of segment spectra was carried out by convolution with a Gaussian of $\sigma = 0.1$ THz and the resulting thinned spectrum was again smoothed by convolution with a Gaussian of $\sigma = 0.5$ THz. Below $\sim$40 THz, the spectrum is very similar to that for SPC/E, but shows peaks for both intramolecular modes at $\sim$50 and $\sim$100 THz. The OH-stretch mode at $\sim$100 THz lacks the collectivity of the DFT MD spectrum, which is mostly due to the interactions of lone-pair Wannier centers with donor hydrogens across HBs.

\section{Error Estimates}
\label{sec:SI_errors}

\begin{figure}[!ht]
	\centering
	\includegraphics{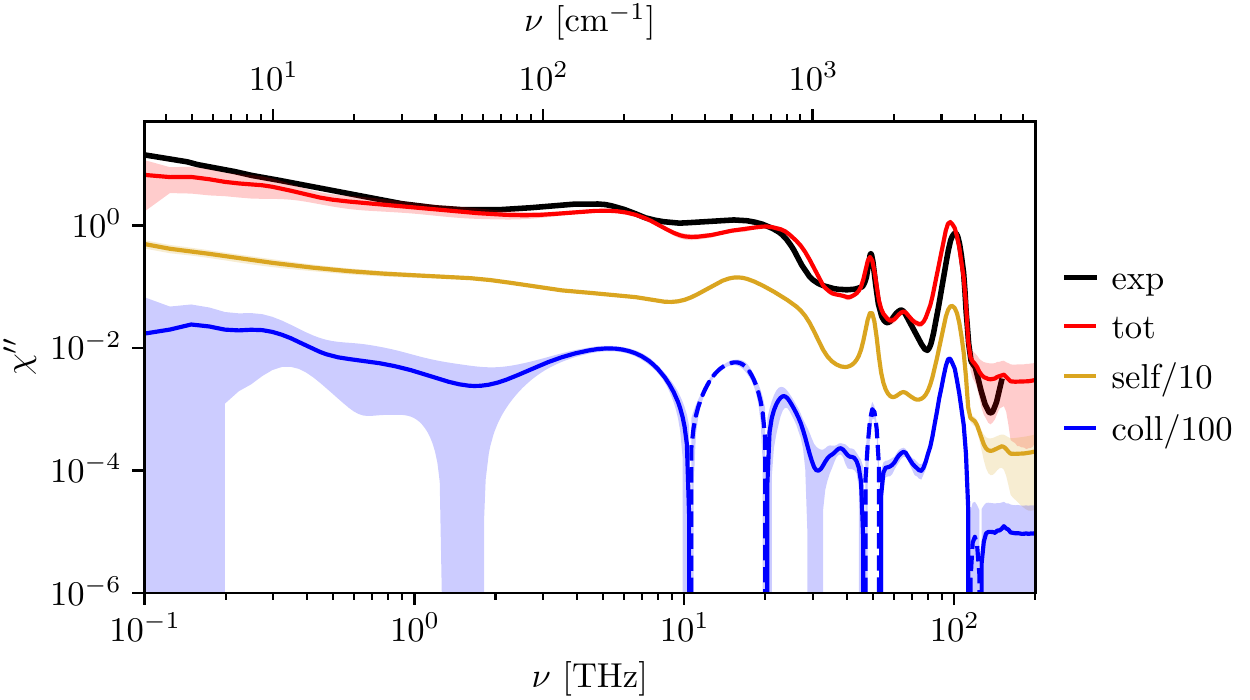}
	\caption{The DFT MD spectrum along with its self and collective components, shifted down by factors of 10 and 100 respectively, shown with their estimated standard deviations (shaded areas).}
	\label{fig:SI_errors}
\end{figure}

Spectra for all figures in this and the main work are calculated as the mean of ten smoothed spectra, from which the standard deviation is obtained. Error estimates are omitted from figures for clarity, excepting one prototypical example, Figure \ref{fig:SI_errors}, where they are shown for the DFT MD spectrum and its self and collective components along with their respective standard deviations, shown as shaded zones above and below the curves. Generally, as in Figure \ref{fig:SI_errors}, standard deviations are small for self spectral components, and large for collective ones.

\section{Acceptor Lone Pair and Donor Hydrogen Interactions}
\label{sec:SI_AWC_DH}

\begin{figure}[!ht]
	\centering
	\includegraphics{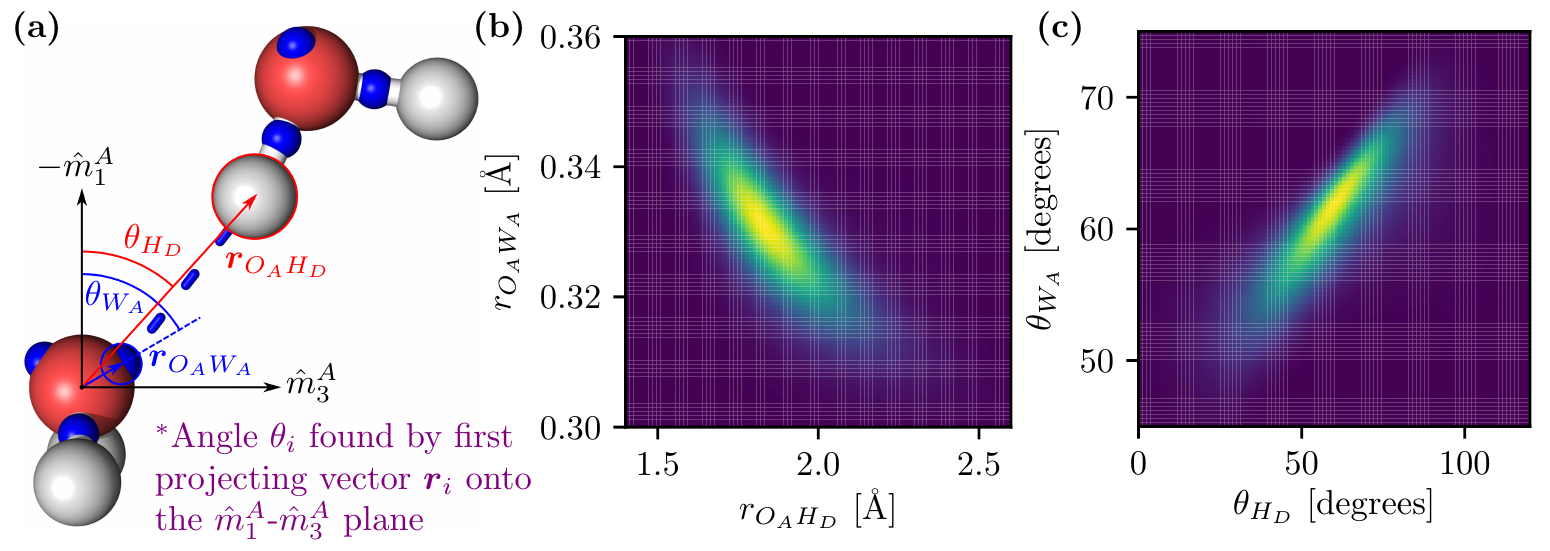}
	\caption{Positional dependence of acceptor Wannier centers on donor hydrogens. \textbf{(a)} Schematic illustrating vector and angle definitions. Wannier centers are represented as small blue spheres. Vectors $\boldsymbol{r}_{O_A H_D}$ and $\boldsymbol{r}_{O_A W_A}$ are the displacement vectors from acceptor oxygen ($O_A$) to donor hydrogen ($H_D$) and the acceptor Wannier center nearest to $H_D$ ($W_A$). Angles $\theta_{H_D}$ and $\theta_{W_A}$ are between the acceptor's orientational axis ($-\hat{m}_1^A$) and the component of $\boldsymbol{r}_{O_A H_D}$ or $\boldsymbol{r}_{O_A W_A}$ in the $\hat{m}_1^A$-$\hat{m}_3^A$ plane. Thus $\theta_{H_D}$ and $\theta_{W_A}$ are independent of position along $\hat{m}_2^A$.  \textbf{(b)} Bivariate histogram over distances $r_{O_A H_D}$ and $r_{O_A W_A}$, which reveals a pronounced negative dependence (yellow represents a high count, navy blue a low count). \textbf{(c)} Bivariate histogram over $\theta_{H_D}$ and $\theta_{W_A}$, which reveals a pronounced positive linear dependence.}
	\label{fig:SI_AWC_DH}
\end{figure}

To aid in understanding the following, refer to the schematic Figure \ref{fig:SI_AWC_DH}a, which shows a snapshot of a hydrogen-bonded dimer from our DFT MD simulation (small blue spheres indicate Wannier centers). We define two water molecules at a simulation timestep as hydrogen-bonded when they fulfill the geometrical Luzar criterion \cite{1996_Luzar_PRL}, and denote the donor hydrogen as $H_D$ and the acceptor oxygen as $O_A$. For our DFT MD trajectory, calculation of Wannier centers and assignment of each to the nearest oxygen consistently results in exactly four Wannier centers per water molecule, arranged around the oxygen in a roughly tetrahedral configuration, with two lying along the OH-bonds, and two on the back side of the oxygen where the lone pair electron density is high. For each hydrogen-bonded dimer in the DFT MD trajectory, we define the lone pair Wannier center of the acceptor molecule that is nearest the donor hydrogen as the ``acceptor Wannier center'', denoting it $W_A$.

We find that an acceptor Wannier center's position with respect to the acceptor oxygen $\boldsymbol{r}_{O_A W_A}$ tends to depend strongly on relative donor hydrogen position $\boldsymbol{r}_{O_A H_D}$, which we interpret as resulting from the Coulomb attraction between donor hydrogens and lone pair electrons. In this picture, lone pair electrons might be expected to stretch away from the parent oxygen as a donor hydrogen approaches. Indeed, Figure \ref{fig:SI_AWC_DH}b agrees with this picture; it shows a bivariate joint histogram over the distances $r_{O_A H_D}$ and $r_{O_A W_A}$, which show a clear negative correlation. As $r_{O_A H_D}$ can be expected to fluctuate under molecular translations (HB stretching) and OH-stretching, this relationship has significant implications for the peaks at $\sim$5 and $\sim$100 THz.

We define the angle between the acceptor's orientational axis ($-\hat{m}_1^A$) and the components of $\boldsymbol{r}_{O_A H_D}$ or $\boldsymbol{r}_{O_A W_A}$ in the $\hat{m}_1^A$-$\hat{m}_3^A$ plane as $\theta_{H_D}$ and $\theta_{W_A}$ respectively.  Figure \ref{fig:SI_AWC_DH}c shows a bivariate histogram of $\theta_{H_D}$ and $\theta_{W_A}$, where there is a positive linear correlation. This indicates that lone pair electrons, attracted by donor hydrogens, track their movement about the acceptor oxygen. The angle $\theta_{H_D}$ can be expected to fluctuate under librations of the acceptor and/or donor molecules, and under bending of $\angle$HOH of the donor molecule, so this relationship has significant implications for the peaks at $\sim$20 and $\sim$50 THz.

\section{Further Studies of Librational Spectra}
\label{sec:SI_librations}

\begin{figure}[!ht]
	\includegraphics{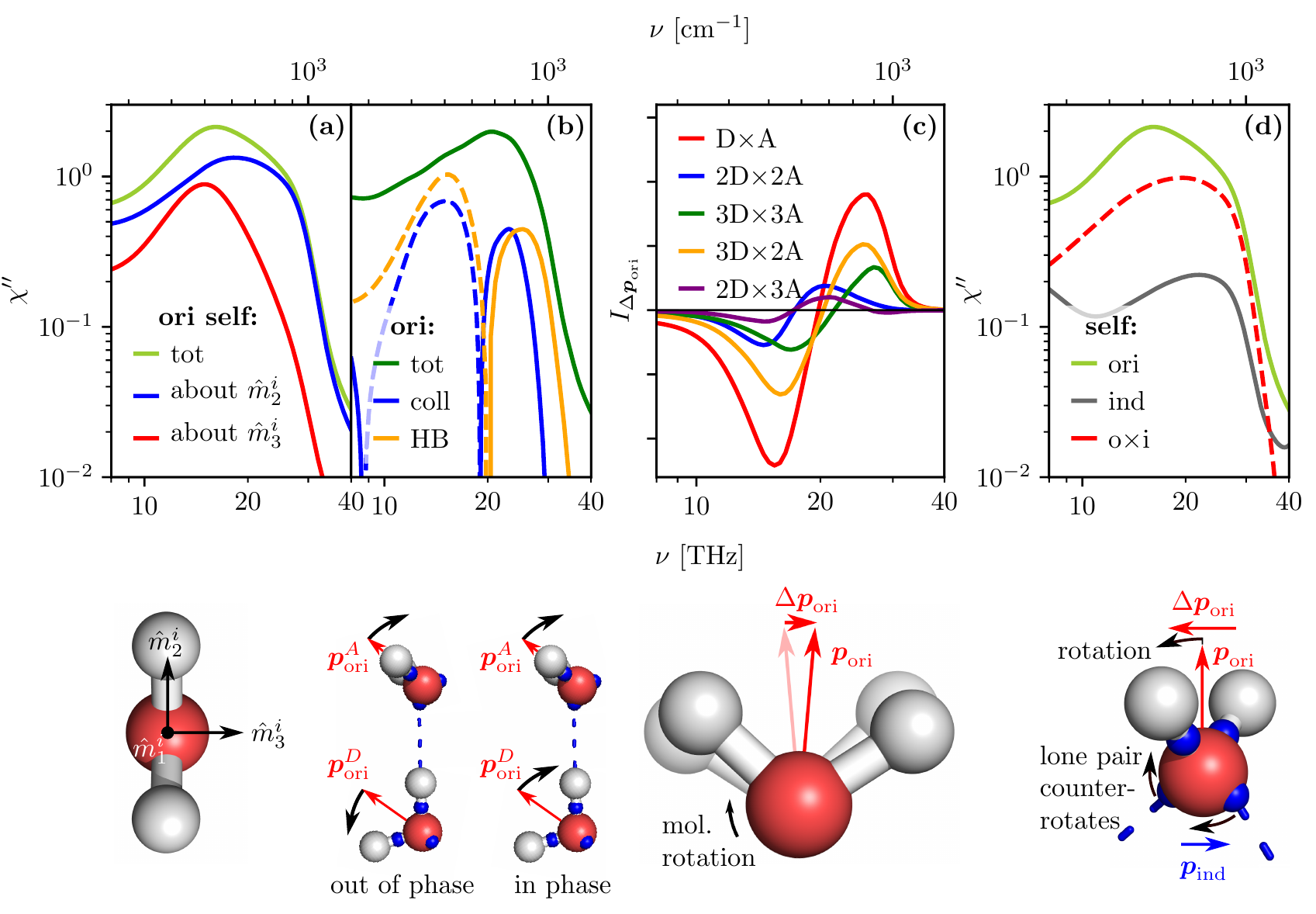}
	\caption{Components of the susceptibility and other power spectra in the libration regime. The schematic below each plot pertains to that plot. Small blue spheres in the schematics are Wannier center positions.
	\textbf{(a)} The orientational self component, along with the subcomponents due to librations about $\hat{m}_2$ and $\hat{m}_3$.
	\textbf{(b)} The orientational spectrum and two of its subcomponents: the collective component and the part of the collective component due only to cross correlations within hydrogen-bonded dimers, which is apparently the source of the in- and out-of-phase modes. 
	\textbf{(c)} The spectral components due to cross correlations in the change in the orientational component $\Delta \boldsymbol{p}_\mathrm{ori}$ over one simulation time step (see schematic) between hydrogen-bonded donors and acceptors. $\Delta \boldsymbol{p}_\mathrm{ori}$ is decomposed into its $\hat{m}_2^i$ and $\hat{m}_3^i$ components for both the donor and the acceptor: the cross spectral components among these are shown in the plot. E.g.\ 2D$\times$3A denotes the cross component due to $ (\hat{m}_2^D \cdot \Delta \boldsymbol{p}_{\mathrm{ori}}^D) \hat{m}_2^D \cdot (\hat{m}_3^A \cdot \Delta \boldsymbol{p}_{\mathrm{ori}}^A) \hat{m}_3^A$. These correspond to librations about different molecular axes. It is clear from the plot that there are out-of-phase and in-phase modes at lower and higher frequencies for all combinations of libration axes.  
	\textbf{(d)} Subcomponents of the self component: namely its orientational, induced, and orientational$\times$induced cross components. The negative cross component implies that molecular dipole moments are induced opposite to librations, the mechanism shown in the schematic.}
	\label{fig:SI_librations}
\end{figure}

In a 1964 study of Raman spectra of water, normal mode analysis of intermolecular motions in a $C_{2v}$-symmetric tetrahedral water cluster predicted modes at roughly 13, 17 and 22 THz, due to librations about $\hat{m}_1$, $\hat{m}_3$, and $\hat{m}_2$ respectively \cite{1964_Walrafen_JCP, inbook_1972_Walrafen_Physics, 1990_Walfren_JPC}. These differences in frequency have been attributed to differences in the moment of inertia of a water molecule about these three principle axes \cite{1990_Walfren_JPC}. As the libration about the $C_{2}$-axis doesn't change the system dipole for the $C_{2v}$-symmetric system, only the 17 and 22 THz modes were predicted to be IR active. A 1995 study fit experimental IR absorption spectra in the libration regime with Gaussians, finding two modes at roughly 11.5 and 20 THz \cite{1995_Zelsmann_JMS}. Figure \ref{fig:SI_librations}a shows the spectral contributions due to librations about $\hat{m}_2$ and $\hat{m}_3$ from our DFT MD simulation. The librations about $\hat{m}_3$ peak sharply at $\sim$15 THz, and those about $\hat{m}_2$ peak at $\sim$18 THz with a shoulder at $\sim$25 THz. This is in relatively good agreement with the normal mode analysis in  Ref.\ \cite{1964_Walrafen_JCP}.

A 2008 study of  IR spectra extracted from Carr-Parrinello simulations of water revealed features at roughly 11.5 and 18 THz in components of a self/collective decomposition, where the collective component was negative at 11.5 THz and positive at 18 THz \cite{2008_Chen_PRB}. These negative and positive modes appear in the collective components of DFT and forcefield MD as well at $\sim$15 and $\sim$25 THz respectively (see main text). Figure \ref{fig:SI_librations}b plots the orientational component, its collective component, and finally a further subcomponent of this collective component: that due only to cross correlations between hydrogen-bonded molecules, which apparently accounts almost entirely for the collective behavior of the orientation there. Thus, we can rule out correlations among non-hydrogen-bonded molecules and conclude that it is in- and out-of-phase dimer modes that underlie these positive and negative collective features, as illustrated in the schematic below Figure \ref{fig:SI_librations}b.

As the frequencies of these in- and out-of-phase dimer modes correspond roughly to those of the different libration directions, a naive explanation is that librations about $\hat{m}_2$ tend to correlate positively between hydrogen-bonded molecules, and those about  $\hat{m}_3$, negatively. Figure \ref{fig:SI_librations}c shows the power spectra defined via $I_x(\nu) \sim | \tilde{x}(\nu) |^2$, in lin-log, due to cross correlations between hydrogen-bonded molecules of $\Delta \boldsymbol{p}_\mathrm{ori}$, the \emph{change} in the orientational component. Further, $\Delta \boldsymbol{p}_\mathrm{ori}$ is decomposed in each molecule into $\hat{m}_2$ and $\hat{m}_3$, (which correspond closely to librations about $\hat{m}_3$ and $\hat{m}_2$ respectively), among which cross power spectra are calculated and plotted. For example, the configuration shown in the schematic below Figure \ref{fig:SI_librations}b is from correlations between $ (\hat{m}_2^D \cdot \Delta \boldsymbol{p}_{\mathrm{ori}}^D) \hat{m}_2^D$ and $ (\hat{m}_3^A \cdot \Delta \boldsymbol{p}_{\mathrm{ori}}^A) \hat{m}_3^A$, which is labeled 2D$\times$3A in Figure \ref{fig:SI_librations}c. It is clear from  Figure \ref{fig:SI_librations}c that for any combination of librational axes, there are negative and positive modes at lower and higher frequencies respectively, which rules out libration axes as an explanation of the frequency difference between the in-phase and out-of-phase libration modes. Our explanation is simply that the shared hydrogen bond is stretched more for in-phase librations, and therefore exerts a stronger restoring force, resulting in a higher frequency, as illustrated in Figures 4a and b in the main text.

\subsection{Lone-Pair Induced Molecular Dipole Moments Under Libration}
\label{sec:SI_librations_lp_ind}

Finally, Figure \ref{fig:SI_librations}d shows all three components of the orientational/induced decomposition of the self component in the libration regime. The line shapes are broadly similar, though the cross component is negative, which indicates that in a librating molecule, a dipole moment is induced opposite to the libration. Figure \ref{fig:SI_AWC_DH}c in Section \ref{sec:SI_AWC_DH} provides the explanation: under librations of an acceptor molecule, its lone-pair electrons track the relative motion of an attractive donor hydrogen, inducing a dipole moment, which is illustrated in the schematic below Figure \ref{fig:SI_librations}d.

\clearpage

\section{Experimental Data}
\label{sec:SI_exp_data}
We compiled experimental spectral data for water at room temperature from existing literature for comparison with our simulated spectra. The compiled experimental dataset consists of data from five different sources for liquid water at or near 300 K, covering different frequency ranges to overlap:

\begin{figure}[!ht]
\begin{minipage}{0.6\linewidth}
	\flushleft
	\includegraphics{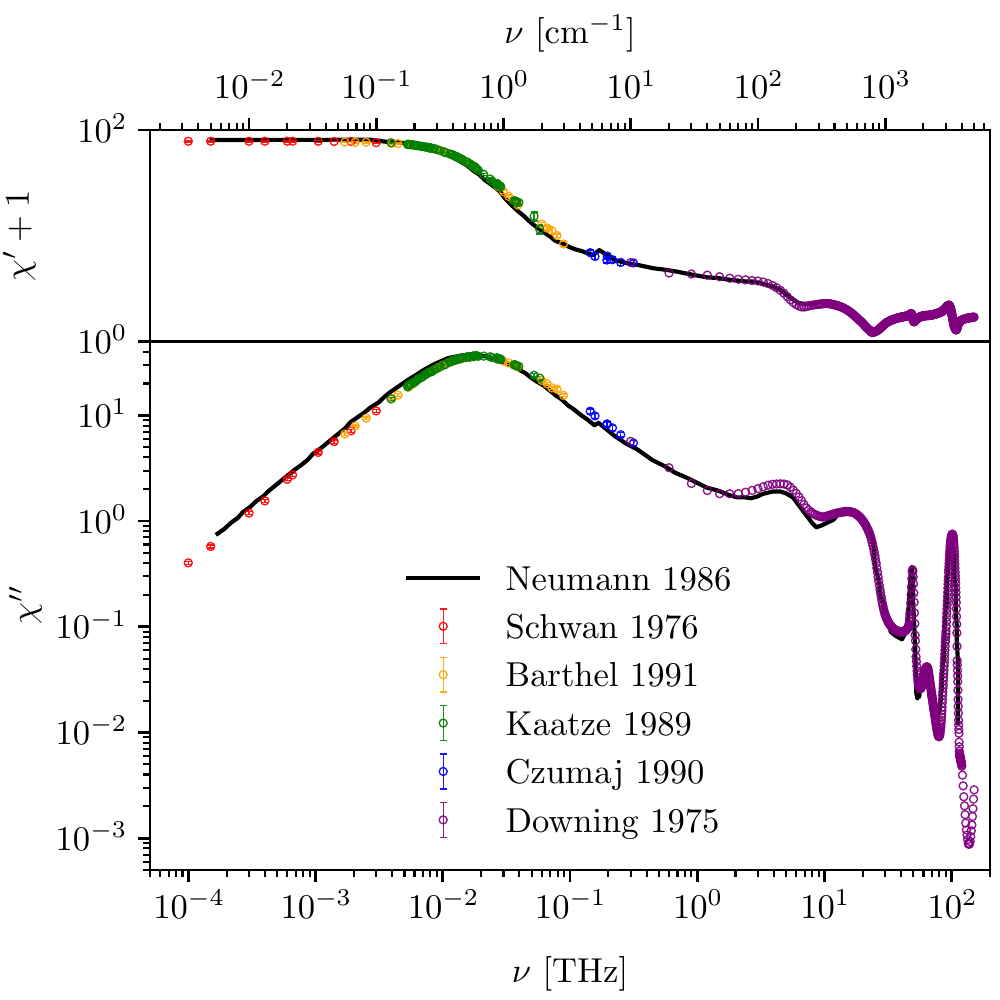}
\end{minipage}
\hspace{5mm}
\begin{minipage}{0.3\linewidth}
	\centering
	\caption{Experimental spectra for water at $300\pm2$ K, compared, with $\varepsilon^{\prime}(\nu)= \chi^{\prime}(\nu)+1$ in the top panel, $\chi^{\prime\prime}(\nu)$ in the bottom. The data shown as colored circles were found in Refs.\ \cite{1976_Schwan_JCP, 1991_Barthel_BBPC, 1989_Kaatze_JCED, 1990_Czumaj_MP, 1975_Downing_JGR}, and are the data that comprise the compiled dataset to be used henceforth for comparison to simulated spectra. The black curve shows another dataset shown only in this figure for comparison \cite{1986_Neumann_JCP}. Error bars are shown for each of the compiled datasets except for the Downing dataset.}
	\label{fig:SI_exp_datasets}
\end{minipage}
\end{figure}
\begin{itemize}
\item Ref.\ 
\cite{1976_Schwan_JCP} 
(Schwan 1976), at a reported temperature of 25$^\circ$C, covers 0.1--3 GHz. It consists of datasets from two separate laboratories, spanning the ranges 0.1--0.7 GHz and 0.8--3 GHz. The highest- and lowest-frequency points of the first and second datasets respectively (at 0.7 and 0.8 GHz) are omitted, as they appear to represent outliers.

\item Ref.\ \cite{1991_Barthel_BBPC} 
(Barthel 1991), at a reported temperature of 25$^\circ$C, covers 1.7--89 GHz.

\item Ref.\ 
\cite{1989_Kaatze_JCED} 
(Kaatze 1989), at a reported temperature of 25$^\circ$C, covers 1.8--58 GHz. The first ten datapoints, covering 1.8--3.75 GHz are noisy, and as this range is also covered by Ref.\ 
\cite{1991_Barthel_BBPC}, they are omitted.

\item Ref.\ 
\cite{1990_Czumaj_MP} 
(Czumaj 1990), at a reported temperature of 25$^\circ$C, covers 57--315 GHz. The first five datapoints, covering 57-90 GHz are noisy, and as this range is also covered by Ref.\ 
\cite{1991_Barthel_BBPC}, 
they are omitted.

\item Ref.\ 
\cite{1975_Downing_JGR} 
(Downing 1975), at a reported temperature of 27$^\circ$C, covers 0.3--150 THz. The quantities given are the indices of refraction and absorption, $n(\nu)$ and $k(\nu)$, from which the susceptibility is calculated via $\chi^{\prime}(\nu)+1 = n(\nu)^2 - k(\nu)^2$, and  $\chi^{\prime\prime}(\nu) = 2n(\nu)k(\nu)$.
\end{itemize}

Thus, the compiled experimental dataset covers a frequency range of 100 MHz to 150 THz. Figure \ref{fig:SI_exp_datasets} shows the compiled data as colored circles compared with a dataset digitized from a figure in Ref.\ \cite{1986_Neumann_JCP} (Neumann 1986) spanning from $\sim$0.2 GHz to $\sim$110 THz.  In figures comparing experimental and simulated spectra in the main work, the data are plotted directly as a heavy black curve without errorbars. Note that Refs.\ 
\cite{1976_Schwan_JCP, 1991_Barthel_BBPC, 1990_Czumaj_MP} 
were found via a compilation in Ref.\ 
\cite{2007_Ellison_JPCRD}.

\clearpage 

\bibliography{spectroscopy}